# Oxide nanotemplates for self-assembling "solid" building blocks

N. Berdunov, G. Mariotto, K. Balakrishnan, S. Murphy, I.V. Shvets

SFI Laboratories, Physics Dept., Trinity College, Dublin 2, Ireland

It is widely accepted that self-assembling building blocks is one of the promising ways for engineering new materials. Recent years reveal substantial progress in fabricating colloidal particles, polymer blocks and supramolecular aggregates of organic molecules. Despite substantial progress in molecular self-assembly there is still a lack of simple blocks made of "solid matter" (e.g. metals, oxides etc.) with well-defined crystal structure and spatial order. Here we demonstrate that ordered arrays of metal nanoclusters can be fabricated by self-assembly on a wide range of oxide templates. These nano-templates are produced either by depositing an alien oxide film or by oxidizing a metal/metal oxide substrate.

Introduction

The future progress of nano-science substantially depends on how successfully we can manipulate the matter down to the atomic scale. Self-assembly of various nano-structures has been widely reported in many branches of science. However, the control of the structure and spatial order, which is essential for spin-transport, quantum-computing and many other applications, remains a major challenge. A template-assisted spontaneous assembly could prove to be a key solution for this problem.

There are spectacular examples of metal clusters self-assembly driven by strain-relief patterns and high energy of step edges [1-2], metal clusters on reconstructed semiconductor surfaces [3] etc. Yet, only a very limited number of such examples are known to this date and the choice of templates is rather scarce.

Metal oxide substrates, ranging from non-conducting to conducting, from non-magnetic to ferro- and antiferro-magnetic, etc., are proved to be reliable for many technological applications. A vast range of reconstructions on metal oxide surfaces makes them very promising for atomic-scale constructing. Furthermore, recent reports show that the difficulty in achieving epitaxial growth on an oxide surface can be overcome [4] and that regular arrays of metal clusters can be formed on nanostructured oxide surfaces [5].

Here we present a wide family of oxide nanotemplates, which provides: a) "set of rules" for self-assembly; b) atomic matching of nanostructures with the substrate (epitaxy); c) tuning of size and spatial order of deposited clusters; d) variations of electron transport scenario between nanostructures and substrate (e.g. conductive/non-conductive, etc).

*FeO/Pt(111) nanotemplate*

The growth of an FeO ultra-thin film on a Pt(111) substrate results in a coincidence structure (Moire pattern) of 2.6 nm periodicity due to the lattice mismatch between film and substrate. In our experiment a single monolayer of FeO grown on Pt(111) (see Experimental) is used as the template for Fe and V deposition. Fe clusters nucleated on the surface are well confined and repeat the surface superlattice (Fig.1a). The regularity of the cluster array improves after a short annealing at 600 K (Fig1.a-inset). The presence of preferential nucleation sites can be seen more clearly in Fig.1c, which shows the formation of vanadium clusters at very low coverage. It should be noted that the appearance of the FeO/Pt(111) coincidence structure in the STM image represents a variation of the local charge density within the surface superstructure rather than a topographical effect [6]. As the film's structural instability (poor matching) leads to a lateral modulation of the surface charge, one can expect a variation of the adsorption properties.

Density Functional Theory (DFT) calculations have been performed to identify the geometry of 1 monolayer (ML) thickness adsorbed metal film. The results suggest that in case of Fe/FeO/Pt(111) the favourable adsorption sites for Fe adatoms are on-top of oxygen. The calculated interlayer distance of 1.8 Å between adsorbed Fe and surface oxygen is well in agreement with STM measured values of 1.5-1.8 Å.

*RhO$_2$/Rh(111) nanotemplate*

Oxide nanotemplates can also be formed by oxidizing the surface of a metal single crystal. As shown in [8], a uniform RhO$_2$ thin film can be stabilized on a Rh(111) substrate . The film represents a O-Rh-O tri-layer forming a coincidence structure with the Rh(111) surface. Although, this phase of rhodium oxide is metastable in the bulk, once it is formed as a two-dimensional oxide layer on top of the rhodium substrate it remains stable at room temperature in a wide range of oxygen pressure. Deposition of an Fe-film on the RhO$_2$/Rh(111) surface results in the formation of an array of iron clusters which, similarly to the Fe/FeO/Pt(111) system, can be improved by annealing in UHV for a short time. An example of a Fe-cluster array formed after depositing of 0.2 Å Fe and post-annealing at 600 K is shown in Fig.1d.

*Oxygen-terminated $Fe_3O_4$ (111) nanotemplate*

The magnetite $Fe_3O_4$ (111) surface reconstructs into a 4.2 nm periodicity superstructure under oxidizing conditions. Our previous research [9] unambiguously indicates that it represents an oxygen-terminated surface, which reconstructs under intrinsic stress creating a long-range order. In this experiment, an $Fe_3O_4$(111) single crystal and a 100nm $Fe_3O_4$(111) film grown on MgO(111) were used. Both substrates were initially annealed in an oxygen atmosphere to form a 4.2 nm superstructure covering almost the entire sample surface (see Experimental for details). Long-range surface order reveals a charge modulation along the surface, as can be seen in bias-dependent STM images (Fig2.a). Deposition of Fe and Cr films results in the formation of regular arrays of nanoclusters on the magnetite surface (Fig.2b,c). In the case of 0.5 Å and 1 Å film thickness, the cluster height is 1 ML [10], whereas a 2 Å thick film results in 2 ML clusters formation conserving the arrays regularity.

**Model and Discussion**

In all three cases presented above, the reason for the long-range surface reconstruction is a structural instability: either intrinsic e.g. due to a non-uniform oxygen lattice as in magnetite [9], or extrinsic, caused by a film's poor matching to the substrate. In some cases, the distortion in the oxygen lattice is evident from STM data [4,9], while in others the existence of the strain in the surface layer can be concluded from the local variations in the stacking sequence. In particular, fcc and hcp local structures, which can be identified in the FeO/Pt(111) coincidence structure (Fig.1b), relax differently resulting in a lateral strain modulation.

To model the effect of the strain in the top surface layer we have performed DFT calculations for the FeO(111) laterally distorted structure. Figure 3 shows charge density difference in the case of two different lattice constants 3.06 Å and 2.8 Å, one is associated with the FeO relaxed on a fcc site and the second on an hcp site (Fig.1b). In both cases, there is significant charge transfer between the surface oxygen and iron in the layer below, which is known to change the surface oxygen properties [11,12]. Besides, as the degree of charge transfer depends on FeO lateral distortion, one can expect the charge modulation across the surface, as observed in the bias-dependent STM images (Fig.2a). The calculated values of adsorption energy for the Fe adlayer gives 1eV stronger adsorption in the case of stretched FeO (3.06 Å lattice). Therefore, one can conclude that the charge reallocation caused by strain (distortion) results in a local change of the surface binding energy within the surface superlattice.

We propose here, that the mechanism of site-selective nucleation is driven by the modulation of strain in the close-packed surface oxygen layer. It works well for oxygen-reactive species and is defined by the oxygen-metal interaction, while for the weak or non-reactive to oxygen metals, the metal-metal interaction controls the growth. This explains the absence of preferential nucleation for Au and Pd clusters growth on FeO/Pt demonstrated in [4,7], when the nucleation process is rather of kinetic mode [1,24].

A close-packed oxygen termination is a common feature of polar surfaces Type 3 for (111) and (0001) orientations of oxide crystals [13], so the other oxides of this family may display similar reconstructions. Among the experimental reports on long-range reconstructions of oxygen-terminated oxide surfaces we identify a number of systems which fall into our nanotemplates family:

| Single crystals and thin films: | Thin film coincidence structures: |
|---|---|
| $Fe_3O_4(111)$[9,14], $Fe_3O_4(111)$/MgO [here], $Fe_2O_3$ (0001)[15], $\alpha\text{-}Al_2O_3$ [16], $FeTiO_3$ (0001)[17], | $RhO_2$/Rh(111)[8], $VO_x$/Rh(111)[18], FeO/Pt(111) [6], FeO/Ru(0001) [19], $TiO_2$/Ru(0001) [20], FeO/Rh(111)[21]; $Al_2O_3$/$Ni_3Al$ [7] |

In these examples the superstructure periodicity ranges from 1.4 nm to 4.5 nm. The future expansion of this list will give us the freedom of tailoring the nanoblocks in size and the array spacing. The different spin-electron transport properties of the substrates provide enormous potential for engineering high-performance magnetic and spin-transport materials.

**Conclusion**

We have shown that strained oxygen terminated oxide surfaces can be used as nanotemplates for fabrication of regular arrays of metal nanoclusters. A large family of metal oxides, which can be used as nanotemplates, provides the possibility of tailoring the self-assembling blocks. Besides, the broad range of magnetic and electronic properties makes these templates attractive for different applications.

**Experimental**

All experiments were performed in UHV conditions at base pressure of $2*10^{-10}$ Torr. Four templates, FeO/Pt(111) and $RhO_2$/Rh(111) thin films, $Fe_3O_4$ single crystal and 100 nm $Fe_3O_4$(111) film grown on MgO(111) substrate (see for growth conditions [22]), have been used in the experiments. FeO film was formed by depositing 1Å of Fe and post oxidising it in $10^{-6}$ Torr $O_2$ at 1000 K for 2 min. The preparation procedure of $RhO_2$/Rh(111) is similar to [8] consisted of oxidising a clean rhodium substrate at $10^{-4}$ Torr oxygen and T = 800 K. $Fe_3O_4$ substrates were oxidised in $10^{-6}$ Torr $O_2$ at 1000 K [9]. In the case of the $Fe_3O_4$/MgO film the annealing time was only 2 min to prevent significant diffusion of Mg to the surface.
Metal thin films were deposited by e-beam evaporation at room temperature using a crystal balance monitor for thickness control. STM results were obtained in constant current mode at room temperature.

**Calculations**

The CASTEP program [23] as a module of Materials Studio was used in our DFT calculations to find a geometry of Fe films grown on the FeO/Pt(111) substrate. A one unit cell vacuum slab similar to that used in [4] has been constructed. The local density LDA functional based on ultrasoft pseudopotentials was used to optimize the surface geometry. A plane wave cutoff of 320eV and a (3x3x1) grid of *k*-points were chosen.

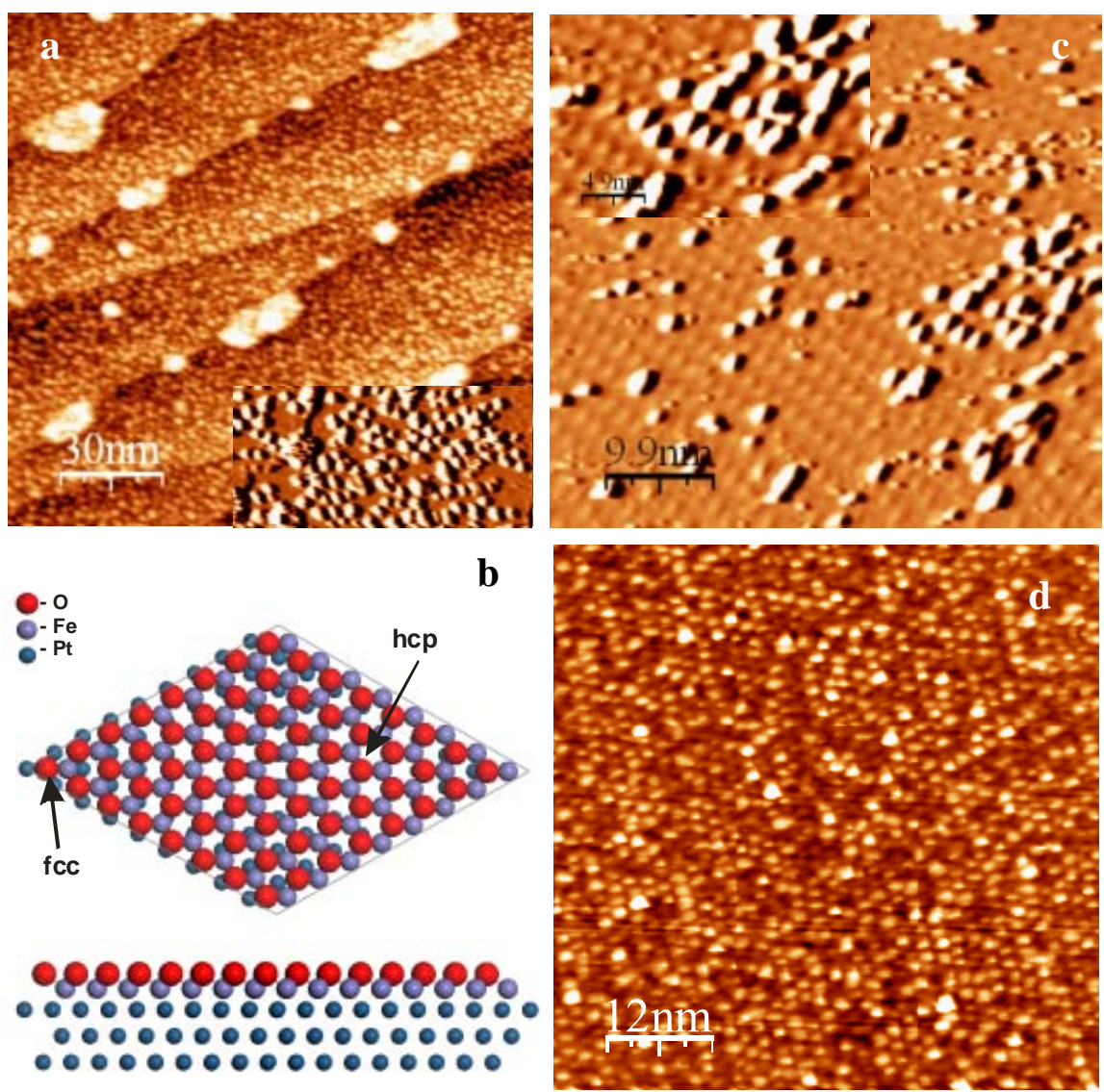

Fig.1. (a) STM image of 0.2 Å Fe film grown on FeO/Pt(111) template. Clusters of 1 ML thickness nucleate following the long-range surface superstructure of 2.4 nm periodicity. Inset shows the same surface (a) after annealing at 600K, which results in more confined clusters' arrangement; (b) Common model of FeO/Pt(111) coincidence structure: FeO bi-layer oxide film stabilized on Pt metal substrate (top-view and side-view); (c) Vanadium clusters nucleate at a particular part of surface superstructure after 0.05 Å vanadium film was deposited (inset shows a higher contras image). Smaller corrugation seen on the STM image due to variations of the local charge density, not the topography effect; (d) Fe clusters formed on $RhO_2$/Rh(111) template after depositing 0.2 Å of iron and annealing at 600K;

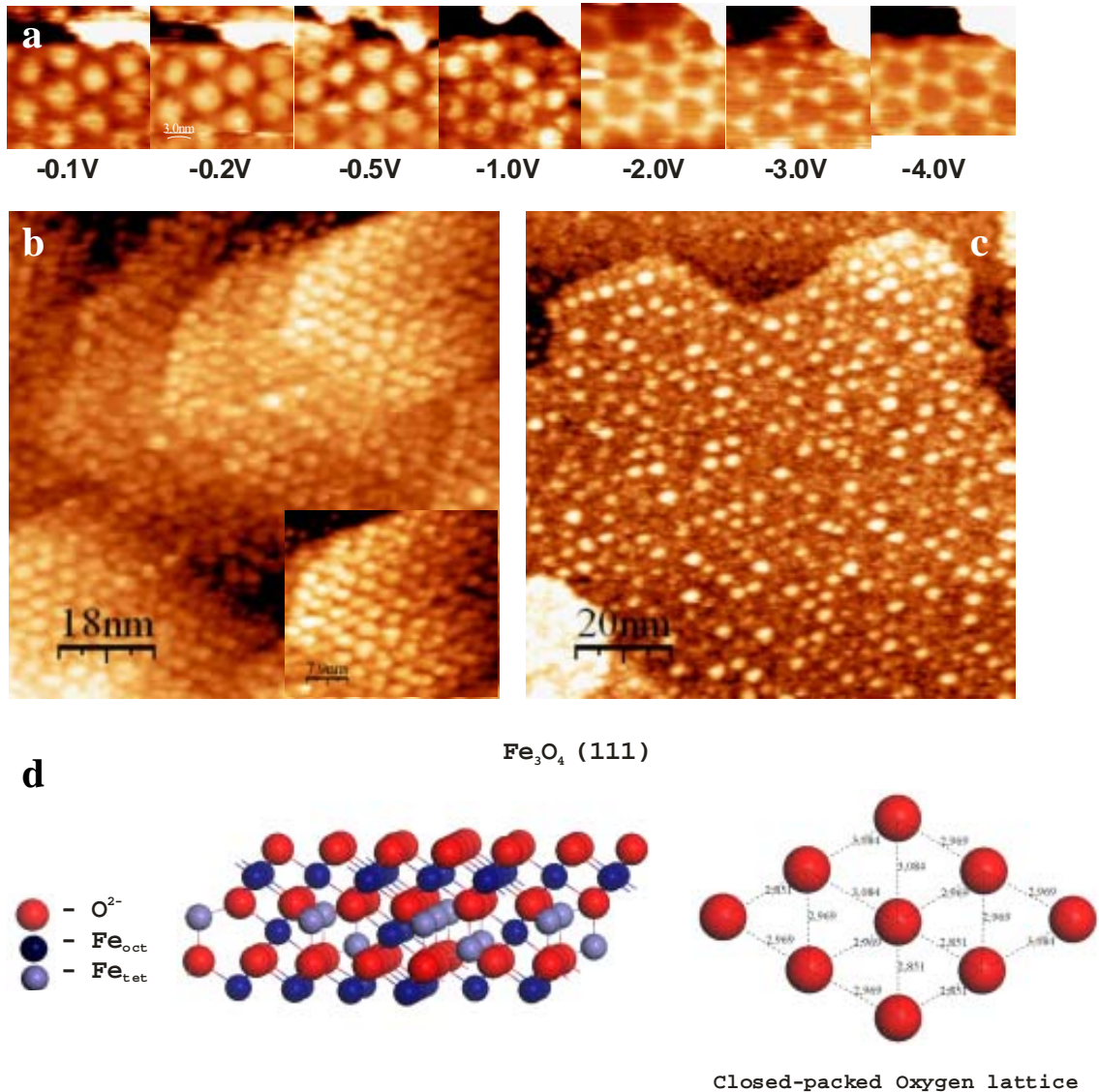

Fig.2. (a) Series of STM images of $Fe_3O_4$(111) superstructured surface taken at different tunneling bias showing the inversion of tunneling contrast; (b) STM images of 2 Å Fe-film deposited on 100nm film $Fe_3O_4$/MgO(111) and (c) 0.5 Å Cr-film deposited on single crystal $Fe_3O_4$(111). Initially, both samples were annealed in oxygen atmosphere to form an oxygen-terminated surface reconstructed to 4.2nm superstructure. Clusters of 2 ML (b) and 1 ML (c) thickness form an ordered array of superstructure periodicity; (inset shows an enhanced contrast image);
(d) Magnetite oxygen-terminated bulk structure (side view) and closed-packed oxygen lattice (top-view).

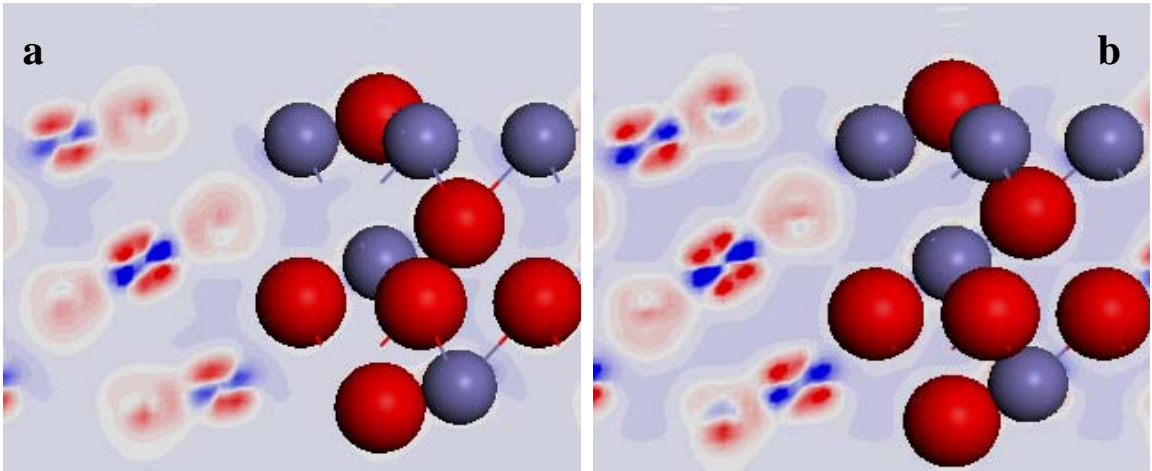

Fig.3. The cross view of FeO vacuum slab and charge density difference distribution for stretched (a) and compressed (b) FeO with corresponding lattice constants of 3.06 Å and 2.8 Å.